# Ni$_3$Si$_2$ Nanowires for Efficient Electron Field Emission and Limitations of the Fowler-Nordheim Model

Amina Belkadi [1], Emma Zeng [2], and A. F. Isakovic [2,a)]

[1] [1]University of Colorado at Boulder, Dept. of ECE Engineering, 425 UCB Boulder, CO, 80309
[2] Colgate University, Dept. of Physics and Astronomy, 13 Oak Dr., Hamilton, NY, 13346

[a)] Electronic mail: aisakovic@colgate.edu and iregx137@gmail.com

The paper reports on top-down nanofabricated Ni$_3$Si$_2$ nanowires and tests of their electron field emission capabilities. The results include low turn on electric field, $E_{ON}$, moderate work function, $\Phi$, and the field enhancement factor, $\beta$, customizable through nanofabrication. The paper also reports on the issues ahead in the field of nanowires-based electron mission, as there are quantitative limitations of the applicability of the Fowler-Nordheim model, which will become increasingly apparent as we continue to optimize field emission of electrons. To this end, we suggest adding the studies of surface-to-volume ratio effects of the nanowires as another standard for comparison, in order to lead to the input form of the density of states as quantum effects becoming more prominent.

## I. INTRODUCTION AND MOTIVATION

The unique nanoscale properties and broad range of electrical, mechanical, and chemical applications make nanowires and the nanodevices comprised of nanowires the focus of many recent studies. The different properties of individual nanowires render their theoretical modeling and experimental characterization a difficult task. Some of the applications of electron emission include, but are not limited to, imaging, X-ray generation, sensors, mass spectrometry, energy conversion and linear accelerators. Often the focus of this research is on identifying materials with relatively low work function ($\Phi$), relatively lower turn-on field ($E_{ON}$), while working with grown and/or nanofabricated high aspect ratio emitters. Understanding FE models in nanowires and nanostructures allows the fabrication of more energy efficient nanodevices, since in some types of electronics devices, FE is the internal electron transfer mechanism[1-3]. This work will focus on the field electron emission (FE) from nanowires.

As a brief reminder, FE, also known as Fowler-Nordheim tunneling, is a form of quantum tunneling where electrons leave negatively biased material (i.e. a cathode) towards the positively biased collector (i.e. an anode), while passing over some controllable short distance in vacuum. Historically, the cathode was made from metals, but nowadays other types of materials are also studied, although standard condensed matter classifications of materials need to be carefully examined when one operates on sufficiently small nanoscale.

Field emission is an emission regime where electrons are emitted from a material due to an externally applied field[4,5]. FE occurs by deep Fowler-Nordheim (FN) tunneling of electrons through an approximate triangular barrier[3,5,6] and is commonly analyzed by FN-type equations. The Fowler-Nordheim model expresses the dependence of the current density ($J$) on the applied electric field ($E$), the work function $\Phi$, and the field enhancement factor ($\beta$), such that[7]

$$J = \frac{AE^2}{\Phi t(y)^2} \exp(-B \frac{\Phi^{3/2}}{E} v(y)) \qquad (1)$$

Here, $\Phi$ is the work function of the material in electron volts (eV), while $A = 1.54 \times 10^{-6} (AV^{-2}eV)$ and $B = 6.83 \times 10^3 (eV^{3/2} V \mu m^{-1})$ are constants, and the functions $v(y) = 0.95 - y^2$ and $t(y)^2 = 1.1$, with $y^2 = e^3 E / 4\pi\varepsilon_0 \Phi^2$, $e$ being the charge of the electron and $\varepsilon_0$ the dielectric constant of vacuum. The applied electric field is often measured in V/μm, and the current density in μA/cm². The Eq. 1 could be simplified and transformed for log-linear analysis of data, by plotting $\log(J/E^2)$ as a function of $1/E$, as is the practice in a number of references.

In this work, we present results of FE studies from Ni$_3$Si$_2$ nanowires and compare them to several other results, adding (i) Ni$_3$Si$_2$ nanowires as another material system of interest, (ii) demonstrating that one can make industrially relevant emitters relying on nanofilm growth and nanolithography, and (iii) introducing an additional standard for performance comparison. Section II presents a discussion of the FN model parameters, their typical values present in the contemporary literature, and how they impact the current density calculations for field electron emission. Section III provides experimental results for the case of Ni$_3$Si$_2$ we contribute. Section IV provides discussion and conclusions.

## II. A BRIEF REVIEW OF FIELD EMISSION AT THE NANOSCALE

The FN model was originally developed for individual bulk metals tips and crystalline solids. Numerous experiments have shown that the FN model is widely applicable to various emitting materials, and it is even commonly used as an approximation to describe FE for non-bulk materials[6,7,8].

In the last two decades we have witnessed a dramatic decrease in the physical size of field emitters. Compared to bulk technologies, nanostructures offer the advantage of faster device turn-on time, compactness, and sustainability[3-8]. Yet, with the continuation of miniaturization, several nano-size related phenomena emerge, such as direct tunneling and quasi-breakdown[3,6,7,8], necessitating the investigation of the applicability of those models developed for bulk to nanoscale devices. Thus, even though the Fowler-Nordheim model of field emission is still used for nanoscale materials and devices, it is of interest to explore its limitations when studying nanomaterials[3].

The way the FN model has been used so far implies that it is applicable and general, having provided decent fits, but according to the literature, it is not always an ideal fit. One way to illustrate the limitations of FN model is to use it as a predictor of the relationship among the parameters the community focuses on in contemporary efforts. To this effect, if one wishes to optimize electron emission from nanomaterials, one often studies the relationship between these three parameters: the turn-on electric field ($E_{ON}$), the work function ($\Phi$), and the field enhancement factor ($\beta$).

Field emission process is highly dependent both on the properties of the material, i.e. the work function, and on the field enhancement factor. The work function is an important parameter that generally varies between 1 to 7 eV for most inorganic semiconductors. The work function of a material determines the energy at which tunneling takes place and thus lower work function (Φ) values offer increased emission efficiency[5,8,9]. The field enhancement factor (β) reflects the degree of field enhancement of the tip shape on a planar surface, and it depends on the emitter geometry, crystal structure, and density of emitted current [6,10]. Materials with higher aspect ratios and sharper edges produce a higher field emission current and consequently, a higher field enhancement factor[9,10].

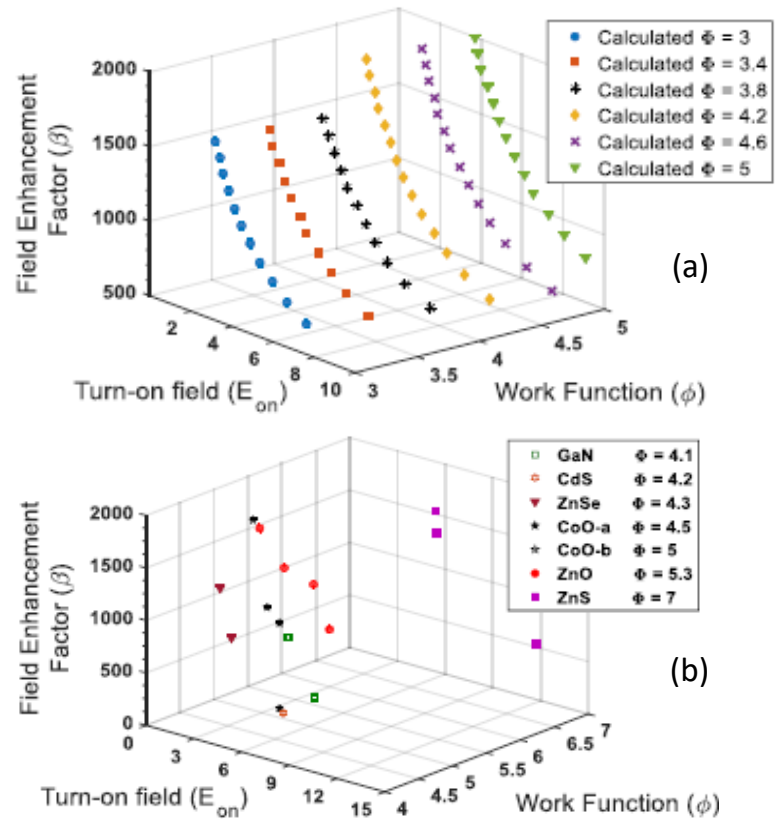

FIG. 1. (a) The relationship between parameters $E_{ON}$, Φ, β based on analytic form of Fowler-Nordheim model predicting smoothly decreasing dependence of the field enhancement factor β for increasing turn-on field $E_{ON}$. (b) Experimentally available values for the same parameters across a broad spectrum of materials used in recent years.

Here, in Fig. 1 (a) we present what the relationship between these parameters *should be* if we are to predict it based on the analytic form of Fowler-Nordheim (FN) model from Eq. 1. The plot is organized so that the work function ($\Phi$) serves as a constant parameter, since it is often assumed that $\Phi$ is material dependent only. We notice a smoothly decreasing value of the field enhancement factor ($\beta$) for increasing turn-on field ($E_{ON}$) at fixed work function values. In a quantitative contrast, Fig. 1(b) represents some of the experimentally available values for the same parameters across a broad spectrum of materials used in recent years, and we see the lack of such "smooth" dependence. There are, of course, more experimentally available values, as the rest of this paper will show.

By carefully checking out the ranges of parameters on the axes between Figs. 1(a) and (b), we see that FN model obtains comparable overall quantitative range of values for all three parameters. The dependence of $\beta$ on $E_{ON}$ is not "smooth", unlike in the prediction. We are aware that this "smooth" relationship from Fig. 1(a) is harder to check without sufficient number of experimental studies on the differently shaped nanowires made from the same material that, due to the design of the materials and that electron emission devices cover a broad range of turn-on field. Nevertheless, the experimentally available points ($E_{ON}$, $\beta$; $\Phi$ = const.) tend to come close to obeying log-normal distribution[11].

An alternative way of looking at these data is to think of the product $\beta E_{ON}$ and examine how relatively constant it is. The standard FN theory uses a constant work function value for bulk, but when it comes to nanostructures, finite size, quantum and surface-to-volume effects suggest that there is some basis to the re-examination of the assumption of a constant work function. One such example is CuO nanowire films, which have an estimated work function varying between 2.5 and 2.8 eV, as obtained from comparison of experimental data with finite element calculation results[12].

The FN theory of FE was based on a smooth infinite plane with a uniform work function, and its applicability to different tip shipped emitters is constantly questioned[12]. The FN theory still applies as a rough approximate of FE for tips with apex radius of at least a few tens of nanometers, which are typically larger than the barrier width. When tips are of the order of 10 nm or less, deviation from FN theory is expected as the apex radius is closer in the value to the width of the potential barrier.

Optimizing nano-field emitters gives space to tune the work function ($\Phi$) and the field enhancement factor ($\beta$), for the sake of enhancing field emission, or achieving the same emission as that obtained from the bulk case but at a lower energy cost. We want to achieve a broad comparison of established, or at least estimated, FE parameters from the model ($\Phi$, $\beta$, $E_{ON}$), and the analytically generated values. FN theory was found to overestimate current densities by a factor of 1.1 and 2.1[3,6,14-17]. The deviation was most prominent for low work functions and high electric fields. Low work functions are the most desirable to work with, and yet that is where deviations are observed.

In our admittedly limited review of the literature of FN model and FE data, electron field emission current density-voltage characteristics were found to only qualitatively follow the conventional FN behavior. Instead, a piece-wise linear behavior appears to provide a better explanation of the results for more than half of studied literature papers[5,6,10,11,18-24]. In addition, several experimental results depict "slight upward curvature" at the lower electric field values for log($J/E^2$) vs (1/$E$) plot, signifying that the Fowler-Nordheim model does not provide a full quantitative characterization of the emission[24-32].

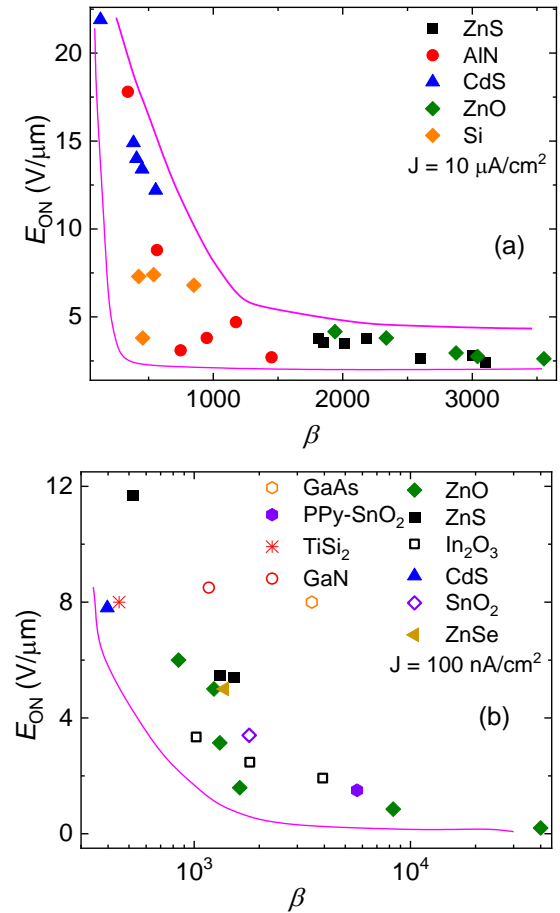

FIG. 2. (a) A short review of the values of the turn on field of some materials as a function of the field enhancement factor $\beta$, for the fixed current density value 10 µA/cm$^2$. (b) same as (a), except for 100 nA/cm$^2$. Note that the symbols are the same in (a) and (b) for materials where nanowires data are available for both current densities. The magenta lines are only a guide to the eye.

Another aspect of verifying the applicability of the FN model is being able to infer interdependencies between the theory's variables. The FN model should be able to predict turn-on fields at different current densities, for particular field enhancement factors and work functions. Turn-on fields are electric fields at which the current densities exponential increase becomes prominent, and that is usually measured at values of 10µm/cm$^2$ or 0.1µm/cm$^2$ for ease and convenience. A survey of measured values of turn-on field and field enhancement factors for different materials and geometries was performed to study the relationship between the turn-on field and the field enhancement factor. Figs. 2(a) and (b) represent the dependence of $E_{ON}$ vs $β$ for current density fixed at 10 µA/cm$^2$ and 0.1 µA/cm$^2$ respectively, for materials with fixed value of Φ between 3.2 and 7 eV. It can be seen from Fig. 2 that although the turn-on field appears to decrease as the field enhancement factor increases (as would be expected by the experimental design, as well), no distinctive, analytically clear trend of that decrease could be identified. In other words, the relationship is not analytical whereas the predicted behavior of $E_{ON}$ vs $β$ from the FN theory is one of inverse proportionality, regardless of the current density at which the turn-on field is taken and regardless of the chosen material work function.

Given which parameters are the focus of this report, we have used two current density values at which we look for the turn on field and the field enhancement dependence, 10 µA/cm$^2$ and 100 nA/cm$^2$, as these values are often used in field emission literature. Two convenient reasons of these choices are: (1) the analysis covers two orders of magnitude incurrent density, with (2) the lower bound being of interest in design of Minimal Energy Electronic Devices (MEES) or Energy Efficient Electronic Devices (E$^3$S). Literature doesn't offer as many results for ($β$, $E_{ON}$) pairs for some of the materials as one might like to "test" for the potential analytical nature of the relationship, but there are enough deviations from any analytical trend that it is safe to say we might want to re-examine the approach based on the "insertion" of the field enhancement term. Of course, there is no reason to expect that all materials would fall onto the exact same curve, but the fact is that the trends are hard to establish. Two additional pieces of information can be gleaned from this type of plot, and we have drawn two lines (in magenta):

(i) It is hard to obtain ($\beta$, $E_{ON}$) point pairs below the curve close to the axes, and

(ii) The point pairs ($\beta$, $E_{ON}$) above the upper magenta line can be obtained but are undesirable (as they would have higher turn on field than desired).

In the recent two decades, efforts on shaping the nanowires' tips and greater variety of materials have contributed to:

(i) More consistent operation at the lower current density

(ii) Lowering of the turn on field, and

(iii) Increase in the $\beta$ parameter.

Regarding (i) we see some materials from the Fig. 2(a) in Fig 2(b), and regarding (iii) we see the need to plot on semi-log scale to obtain a more clear, meaningful comparison.

Conventionally, failure of the FN model in describing field emission is compensated for using the modified FN theory. Dyke and Dolan theory includes the effect of vacuum space charge surrounding the emission site[3,6,14,33]. The FN theory linearity fails at high current densities, due to temperature effects[34,35], among other reasons [35-41]. For semiconductor carbon nanotubes, the breakpoints observed in field emission characteristics is attributed to space-charge build up and carrier saturation[6,7,8,33]. At low field, the space-charge effect should be negligible and at high fields, the space charge and low carrier density should drive the FN plot to saturation, yet that is not observed. Rather, an exponential smooth rise is more prominent. Therefore, this rules out those factors as explanations for breakpoints (piece wise behavior) in the FN plot. Earlier deviations from the FN model were noticed and attributed to the semiconductor nature of the materials, since the FN model was derived for bulk metallic materials[1,3,6]. The deviation was noticed at certain field ranges, typically the low fields. However brief and incomplete our review is, we agree with a majority of the authors who point out that a number of limitations exist in the FN model.

## III. EXPERIMENTAL APPROACH AND RESULTS

We grew $Ni_3Si_2$ thin films (2-5 μm thickness) using sputtering in $10^{-8}$ Torr vacuum. A random circular spot mask pattern was used to perform initial masking for etching with the initial size of the circular spots on the mask in the range 50 – 250 nm. The mask is such that it allows for the control of the density of the nanofabricated nanowires, which is one of the design requests from the applied nanodevices side. Dry etching and ion milling processes were used to remove much of the film, and left $Ni_3Si_2$ nanowires sticking up from the substrate level. The stoichiometric ratio was controlled through the control of sputtering parameters and checked using micro-X-ray fluorescence (micro-XRF).

During the etching process, etch parameters ($Ar^+$, $CF_4$ flux, RF power and sequence timing) were cyclically tuned so to create under-etch profiles in the top part of the nanowires. Naturally, not all nanowires are smooth, so we have chosen to show relative "extremes" – a rough cylindrical shaped nanowire in Fig. 4(b) compared to conically shaped nanowire next to it, which is representative of the more smooth shapes we were able to obtain for some nanowires. At the time of this report writing, the yield of the etch process is still under 70%. Standard field emission current measurements were conducted in a slightly weaker vacuum, but we periodically checked the sample for the presence of oxides on the surface.

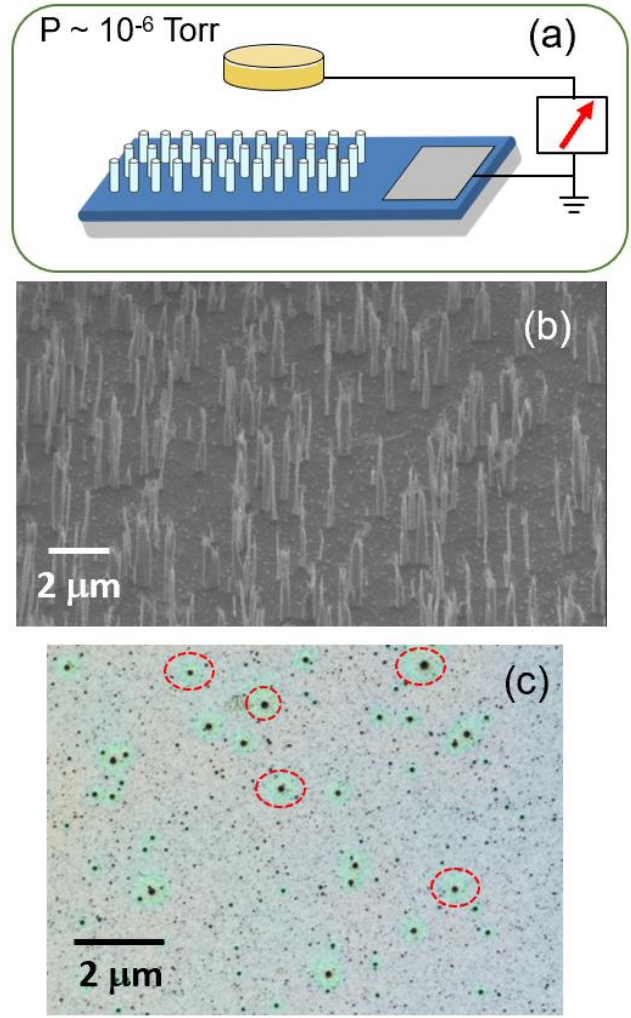

FIG. 3. (a) A simplified sketch of the experimental setup. (b) SEM image of the $Ni_3Si_2$ nanowires pointing up from the surface of the wafer. Nanowires are fabricated by the top down etch process after a random spot mask, which allows for control of nanowires density, is applied. (c) a false color top-down image where some of the nanowires are labeled with red dotted ovals. Green tinted areas around wires are consequence of uneven etching process near wires and correspond to "corrugated" surface visible in the vicinity of the nanowires in panel (b).

Nanowires characterization is presented in Fig. 4. We see that the different shape of $Ni_3Si_2$ nanowires resulted in different transport data, and consequently in the modifications of the turn on field and the enhancement factor. The work function values are 4.3 +/- 0.2 eV, which is among the lowest values reported for the nickel silicides. We note these values make it promising to further study various $Ni_xSi_y$ stoichiometries, with the goal of identifying the lowest possible work function.

The panel (a) in Fig. 4 shows the two relative extremes (most curves collected are to the right of the blue curve and to the left of the red curve). Relatedly, panels (b) and (c) show the relative representative extremes of the nanowires design in our approach. Note that these are "side view" SEM images, so, certain amount of charging effects was unavoidable. The conical shape nanowire points towards future improvements. To this end, we point out that one doesn't have to make a specific shape tip (like "pure cone" as in some references). Instead, what can be done is successive narrowing down (from say 100 nm nanowire diameter, to 50 nm, then to 10 nm). Such approach is doable with controllable nanofabrication in a more or less, standard top-down approach, unlike bottom-up approach in a number of references in the last 20 odd years.

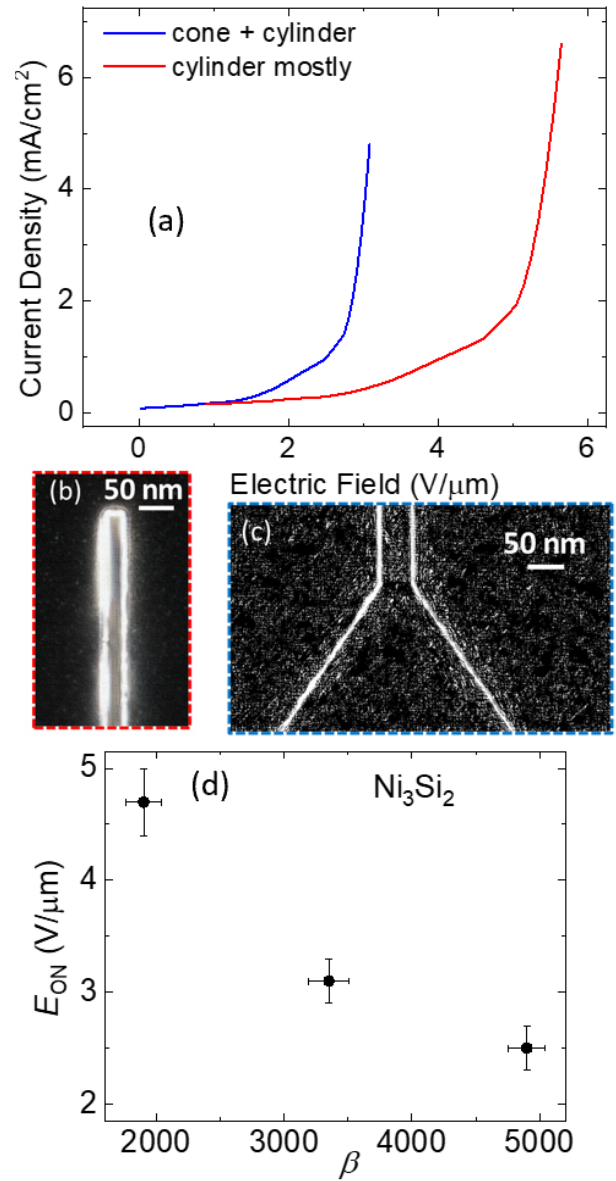

FIG. 4. (a) examples of current density – electric field (J-E) curves for different shapes of the nanowires. (b) side SEM image of the nanowires emitter of predominantly cylindrical shape. (c) side SEM image of the top portion of a nanowire etched in the shape of top-down "funnel", with the goal of modifying its field emission performance. (d) The turn-on field dependence on the parameter $\beta$.

The values for $E_{ON}$ and $\beta$ in Fig. 4(d) are the result of the analysis based on Eq.1. One can alternatively perform the analysis proposed by Forbes and Deane[33]. In our use of Forbes-Deane recommendation, we have decided to vary the exponent in $\log(J/E^n)$ vs $1/E$, instead of keeping it fixed (1.2 value in[33]). The output of the $n$ value lading to the best fit is shown in Table I, with the emphasis on the varied nature of the $Ni_3Si_2$ wires we nanofabricated.

TABLE I – variation of the exponent in Forbes-Deane analysis of $Ni_3Si_2$ emitters

| Wire | Parameter n | Comment |
|---|---|---|
| $Ni_3Si_2$ - A | 1.18 +/- 0.05 | cylindrical tip |
| $Ni_3Si_2$ – B | 1.25 +/- 0.06 | tapered tip (cylindrical to conical) |
| $Ni_3Si_2$ - C | 1.34 +/- 0.05 | conical tip |

There can be additional corrections to analysis of emission current, for example one motivated by[32].

As we make increasingly anisotropic and overall smaller nanowires, there will be an eventual need to re-derive FN model for 2D and 1D DoS, which will likely lead to better quality fitting of the experimental data. Further improvement of these results are possible if accounting for oxidation defects and related surface physics and chemistry phenomena using techniques such as synchrotron based nanofocused X-rays XPS spectroscopy.

Fig. 5(a) presents three images of nanowire during $J$ vs $E$ data taking. Images from the left show several bright spots, and, as we continue to run the same $J$ vs $E$ curves there are fewer such spots, until they completely disappear around 300[th] cycle. After this, the spots never re-appear. Given the scale size, we hypothesize these spots are migrating Ni clusters, or $Ni_xSi_y$, where x > 3 and y < 2. To test the hypothesis, we collected and integrated Auger spectra on a fixed spot on a wire. Fig. 5(b) presents data indicating that the microprobe finds small variations, shown in dashed rectangle) in stoichiometry until 300$^{th}$ cycle is reached.

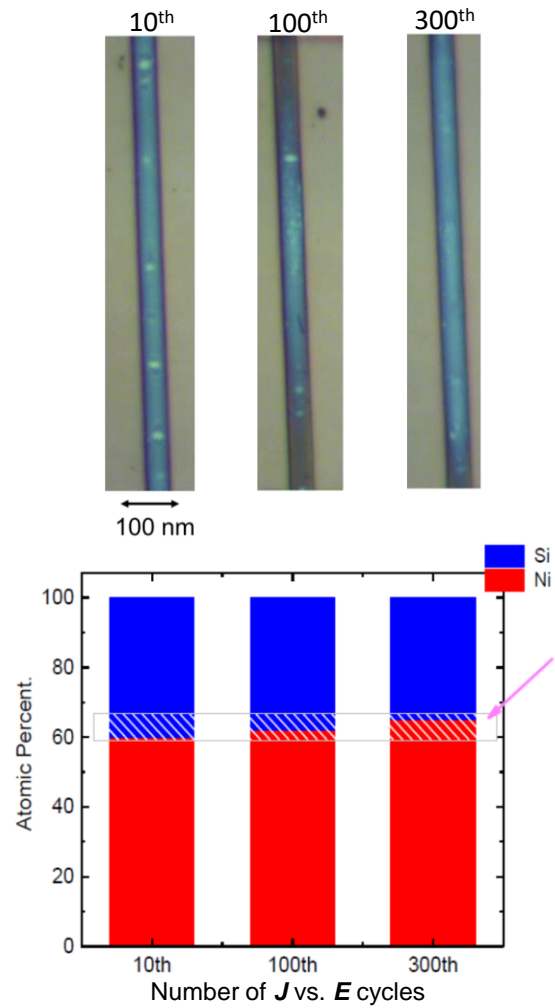

FIG. 5. (upper) three images of the nanowire during ***J*** vs ***E*** data taking. The bright spots seen in the first image decrease in number and then disappear after 300$^{th}$ cycle. The spots are most likely migrating Ni clusters, or $Ni_xSi_y$, where x > 3 and y < 2. Images are obtained with polarization sensitive, sub-micron microscopy setup. (lower) Data from Auger probe microscopy at a fixed spot indicate that microprobe finds small variations, shown in dashed rectangle, in stoichiometry until 300$^{th}$ cycle is reached, after which no variations are observed.

Parameter $\beta$ received a fair bit of the attention in this report. We would like to propose a new measure against which the progress in nanowire field emission is measured. As we make increasingly high aspect ratio nanowires, the surface-to-volume ratio (S/V, hereafter) is increasing. By definition, $\beta = h/r$, where $h$ is the height (or length) of the emitter and $r$ is the equivalent radius of the curvature of the tip. While the dimensionless nature of this parameter makes it appealing for inclusion into the modified Fowler-Nordheim expression currently used, we argue that the surface-to-volume ratio is physically more relevant and captures a good geometric component of the density of states (DoS). Given the variety of parameters in this problem, such as material choices, experimental conditions, geometry, etc., estimating and tracking the systematic progress is a bit of a challenge. Additionally, limited validity of the Fowler-Nordheim model has been a challenge.

We propose a measure of surface-to-volume ratio, S/V, often used in general design considerations in nanoscience and nanotechnology. To illustrate a possibly valid point about S/V ratio, we plotted the values of field enhancement factor, $\beta$, as a function of S/V ratio in two different ways. Before we proceed, we wish to emphasize this was possible only for a subset of all papers available, as we needed clear and relatively unambiguous SEM or TEM images and related scale information, from which we can calculate, or at least, estimate S/V ratio. In one plot, in Fig. 6(a), we calculated S/V ratio for the *entire* nanowire, and in the related plot in Fig. 6(b), we plotted *only the nanowire tip* value of the S/V ratio. It is pretty clear that the two plots offer qualitatively different information. We suggest this in Fig. 6(a), with guide-to-the-eye blue lines which seems to offer boundaries for most nanowires in $\beta$ vs S/V plot. Unlike that plot, Fig. 6(b) shows how most materials cluster in a narrow region of S/V values.

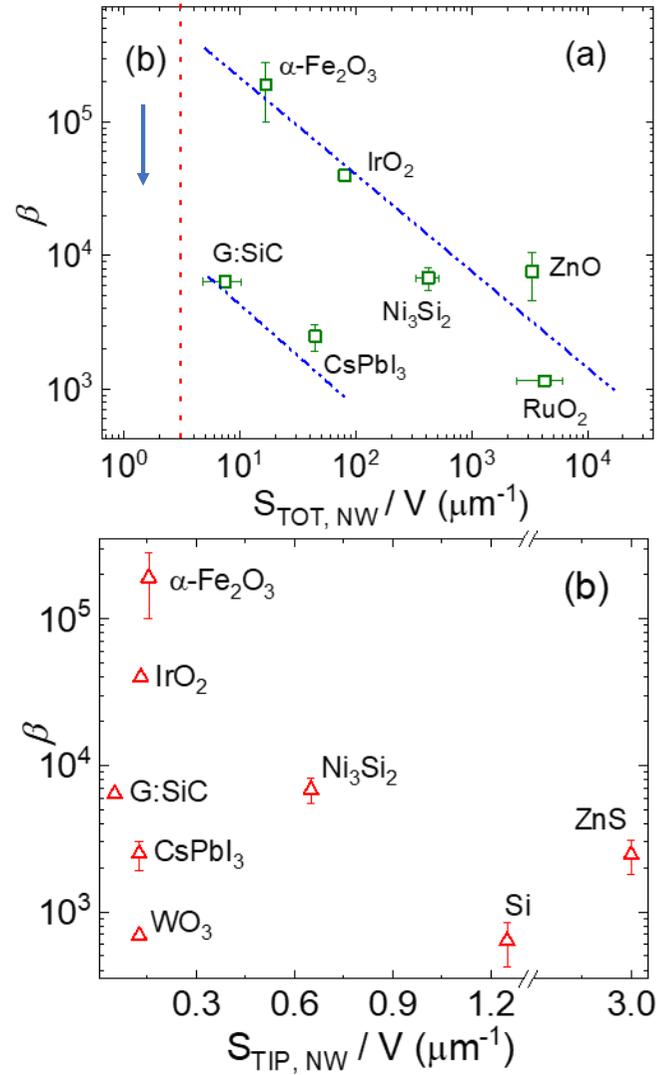

FIG. 6. (a) Field enhancement factor as a function of the total surface-to-volume of the nanowires with possible boundaries indicated as guide-to-the-eye. (b) same plot except only for the tip of the nanowire surface-to-volume values. All values needed for S/V ratio are taken from the references cited here, in addition to our own $Ni_3Si_2$ samples. Note that the range of horizontal axis in (b), coincides with the small portion of the horizontal range on the far left of the upper panel, (a).

# IV. SUMMARY, CONCLUSIONS, and FUTURE WORK

In summary, this manuscript reports on:

(a) the parameter $\beta$ for $Ni_3Si_2$ is subject to nanowire design and nanofabrication changes, in overall agreement with the trends from existing literature on nanowires.

(b) the agreement with the trends for other materials is noticeable because it is possible to obtain lower turn-on field $E_{ON}$, which is among the goals of modern field emission nanodevices for some applications.

(c) there are limits to the applicability of FN model, as indicated in our literature review.

(d) suggested use of S/V ratio to additionally evaluate the performance of nanowire emitters.

It is clear that the trend of increasing S/V ratio calls for the use of non-3D density of states (DoS) in the expression for the current density dependence on the electric field. We hope to add the of S/V ratio as another parameter with which emitter are evaluated because of its value in emphasizing the role of emitters' geometry as researchers continue to make emitter ever sharper and smaller.

Nanowires barely satisfy the criterion implicitly assumed in derivation of the FN model namely because they are not true bulk objects, so one needs to account for the non-3D density of states during the derivation, which is a subject of an ongoing work. It is known that, depending on the material and device system parameters, FE could be either described by FN or Child-Langmuir law in bulk. Instead of a clear separation of these two regimes in bulk, the ongoing work, along the lines of[29-43] offers hope of a novel model. We are conducting our own model development which would take into account fractional density of states. it is worth investigating whether there is a transitional region relevant to NWs that lays between these two clear regimes. Additional efforts are needed to understand nanoscale and atomic scale changes in the nanowires, specifically the role of defects and oxidation. Lastly, the FN model assumes a single emitter and is independent of the surrounding emitters, so any update should be modified to include the interaction between the nanowires[31-32].

## ACKNOWLEDGMENTS

ABB and AFI gratefully acknowledge support from MEES-I SRC program 2011-KJ-2190, while they were both affiliated with Khalifa University, UAE. EZ is grateful for the Colgate University Summer Research support for undergrads. Parts of this work were performed at Cornell Center for Nanofabrication, supported by National Science Foundation, and at Brookhaven National Laboratory, supported by US Dept. of Energy. AFI is thankful to Prof. Richard G. Forbes, for early stimulating exchanges and recent clarifications on the topic of FN limitations. The authors thank the anonymous reviewers whose comments led to the correction of several errors and omissions in the initial manuscript.

## DATA AVAILABILITY

The data that support the findings of this study are available from the corresponding author upon reasonable request.

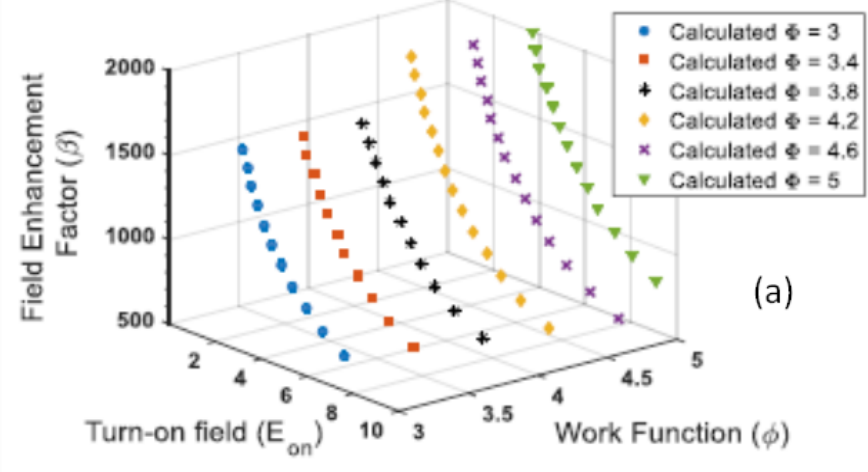

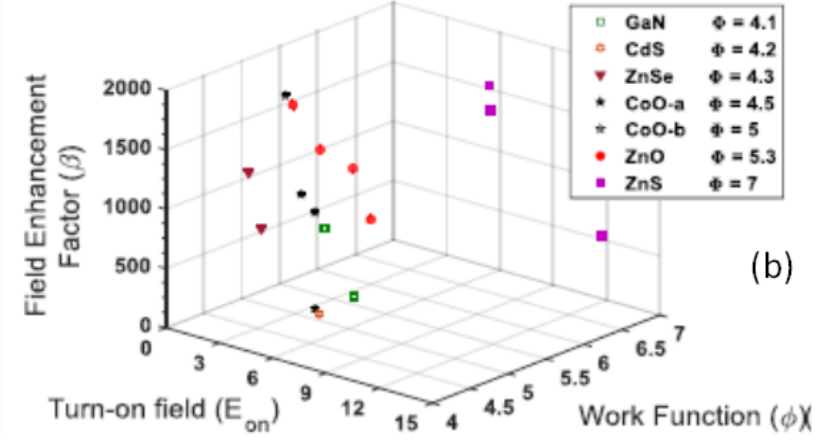

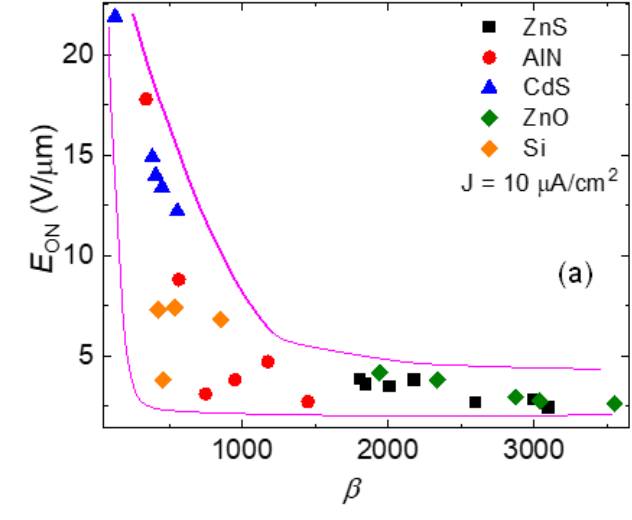

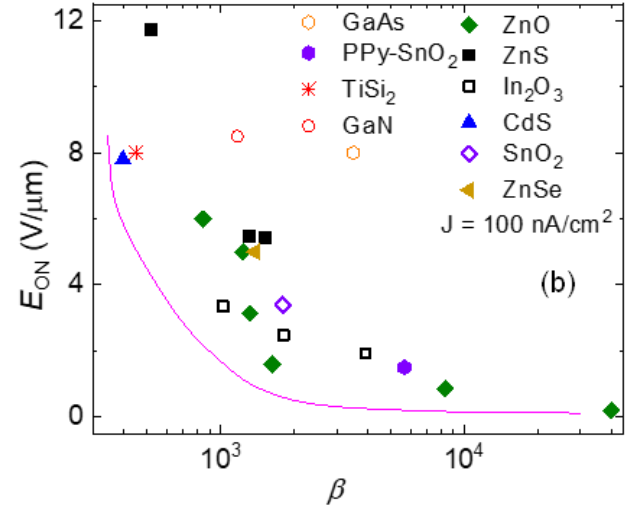

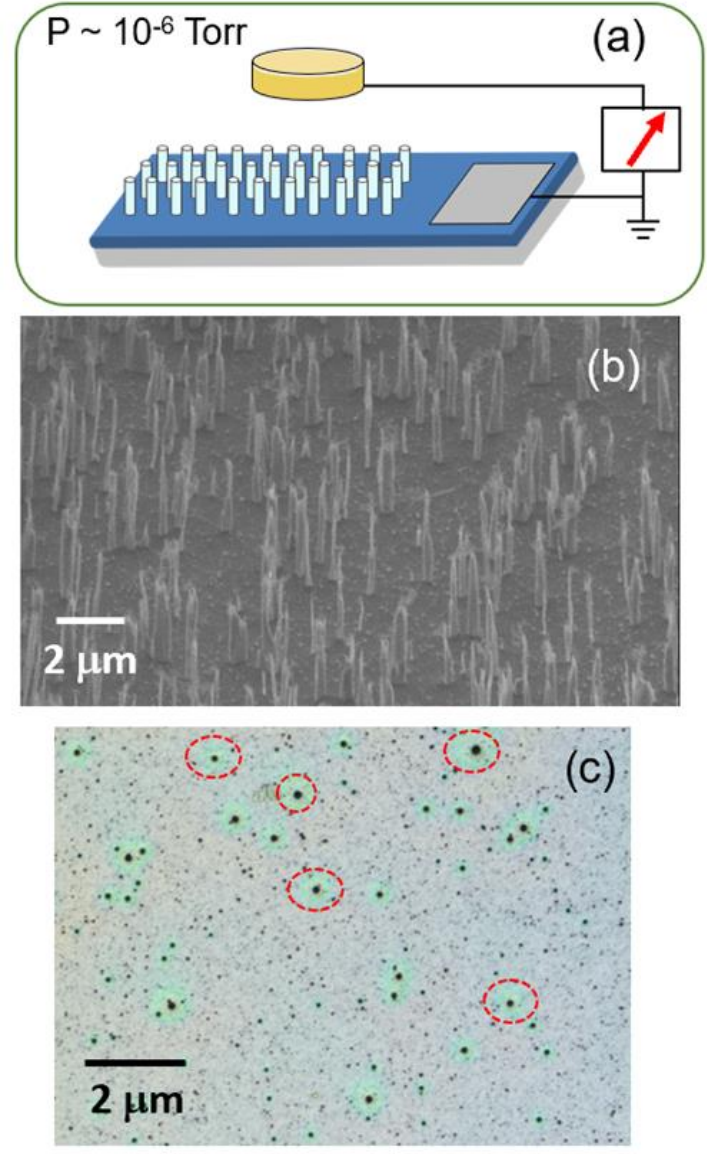

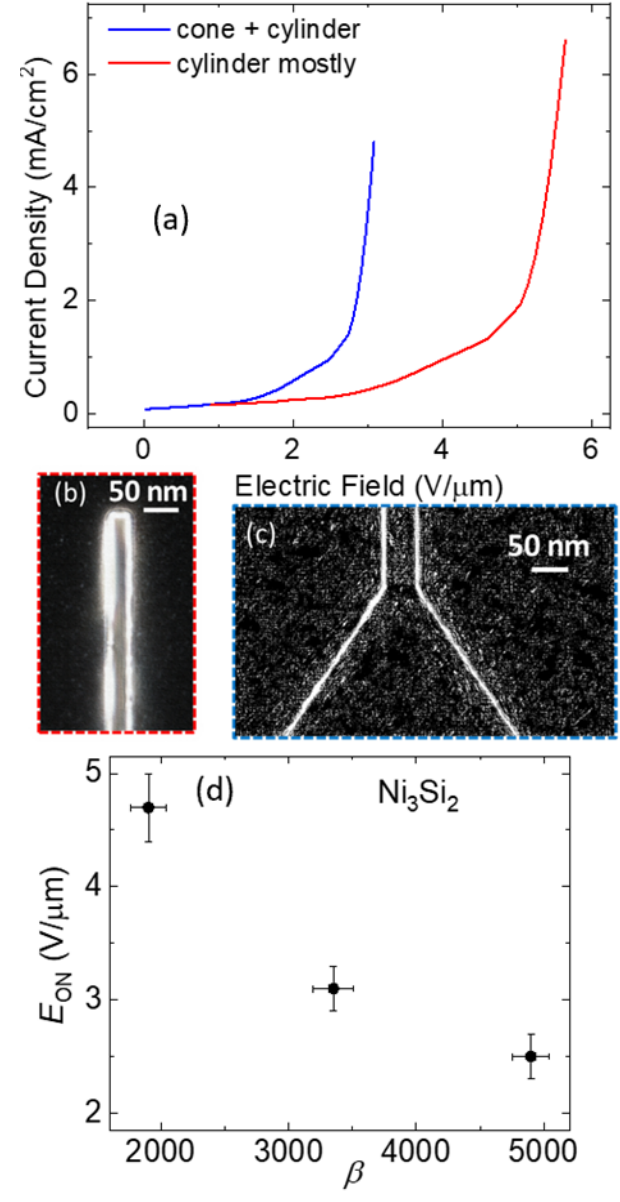

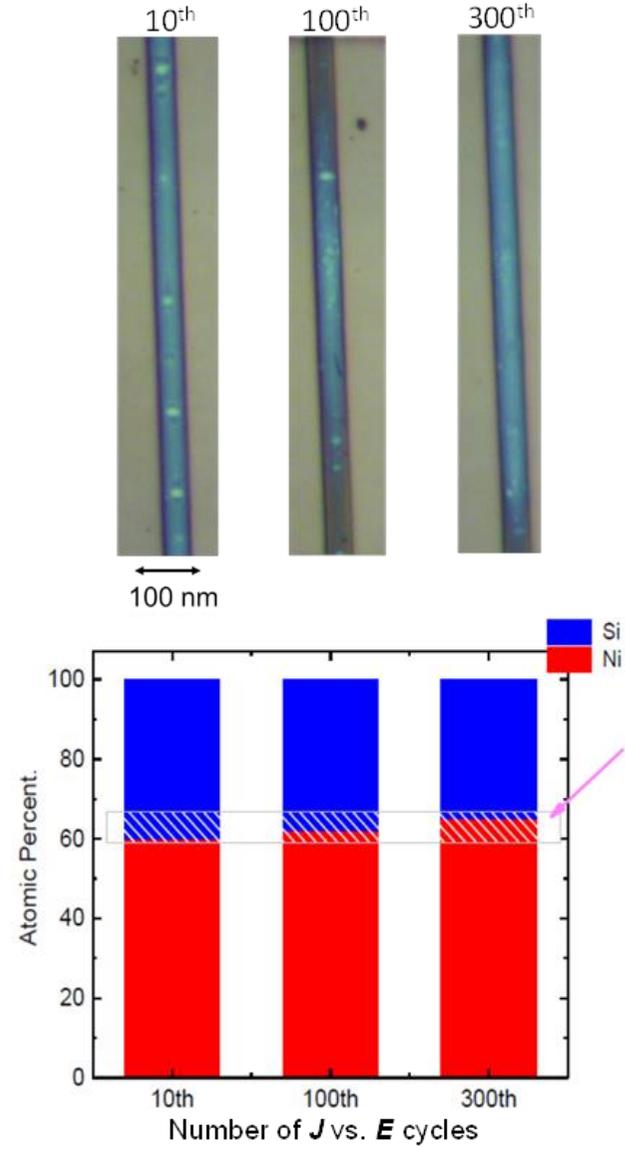

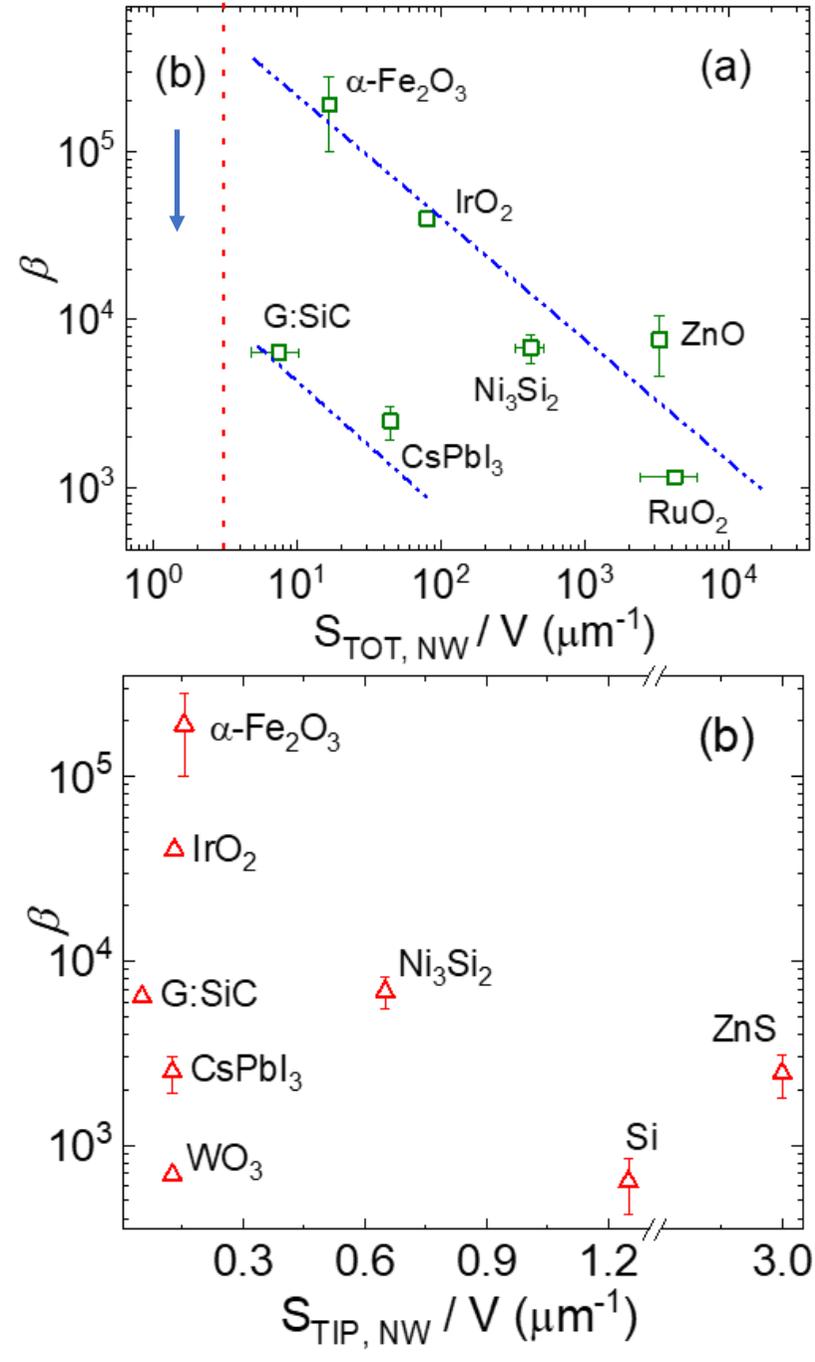